\documentclass[12pt]{iopart}
\usepackage{iopams}
\usepackage{graphicx}
\usepackage{bm}  
\eqnobysec

\begin{document}

\title[Closed form solution for a double quantum well using Gr{\"o}bner basis]{Closed form solution for a double quantum well using Gr{\"o}bner basis}

\author{A Acus\dag, A Dargys\ddag\footnote[3]{Corresponding
author: dargys@pfi.lt}}

\address{\dag\ Vilnius University, Institute of Theoretical Physics and Astronomy,
A.~Go\v{s}tauto 12, LT-01108 Vilnius, Lithuania\\
Vilnius Pedagogical University, Studentu 39, LT-08106 Vilnius,
Lithuania}

\address{\ddag\ Center for Physical Sciences and Technology, Semiconductor Physics
Institute, A.~Go\v{s}tauto 11, LT-01108 Vilnius, Lithuania}
\ead{dargys@pfi.lt}

\date{5 January 2011}

\begin{abstract}
Analytical expressions for spectrum, eigenfunctions and dipole
matrix elements of a square double quantum well (DQW) are
presented for a general case when the potential in different
regions of the DQW has different heights and effective masses are
different. This was achieved by Gr{\"o}bner basis algorithm which
allows to disentangle the resulting coupled polynomials without
explicitly solving the transcendental eigenvalue equation.
\end{abstract}

\pacs{73.20.P, 73.20.D, 78.66, 03.65.G } \vspace{2pc}

\noindent{\it Keywords}: Heterostructures, Double quantum well,
Eigenfunctions, Gr{\"o}bner basis, Optical dipole matrix element, Closed form solution

\submitto{\PS}

\maketitle

\section{\label{sec:1}Introduction}
The square double quantum well (DQW) often is used as a toy model
to demonstrate the interaction between quantized energy levels due
to particle tunneling through a potential barrier separating
individual
wells~\cite{Goldman61,deutchman71,deutchman74,Johnson82,demenezes85,Peacock06}.
Recently the DQW model had attracted considerable attention in
semiconductor heterostructure physics because of its applications
in nanoelectronics~\cite{Weisbuch87,Harrison05,Manasreh05}. The
tunneling conductance properties of semiconducting DQW devices as
well as drag effects that result from interaction between
electrons moving at different velocities in different wells was
recently  discussed, for example, in review
articles~\cite{Hasbun02,Debray02}.

Appearance of transcendental equations that describe DQW spectrum
limits direct application of analytical methods in tackling the
eigenfunction problems. Initially the problem of finding the
eigenfunctions has been solved by perturbation theory assuming
that energy level splitting due to tunneling is
small~\cite{Goldman61}. The most recent analytical approach
heavily relies on symmetry properties of the DQW~\cite{Peacock06}.
Of course, this restriction can be relaxed by resorting to
numerical
methods~\cite{deutchman71,Johnson82,Peacock06,Harrison05,Bastard84}.
However, in many cases a knowledge of analytical form of the wave
function is more preferable. For example, in the wave packet
dynamics problems the closed form solution allows one to construct
a direct superposition of eigenfunctions to make a computational
task easy. Here we demonstrate that one can push the problem
further and calculate the relevant eigenfunctions exactly by
exploiting a computer based Gr{\"o}bner basis
algorithm~\cite{Cox98}. In sections~\ref{sec:2} and~\ref{sec:3}
the spectrum and eigenfunctions of a general DQW are calculated
using the Gr{\"o}bner basis, and in section~\ref{sec:4} the
results are applied to find closed form expression for optical
dipole matrix element of the DQW.

\section{Spectrum\label{sec:2}}
\begin{figure}[t]
\centering
\includegraphics[width=10cm]{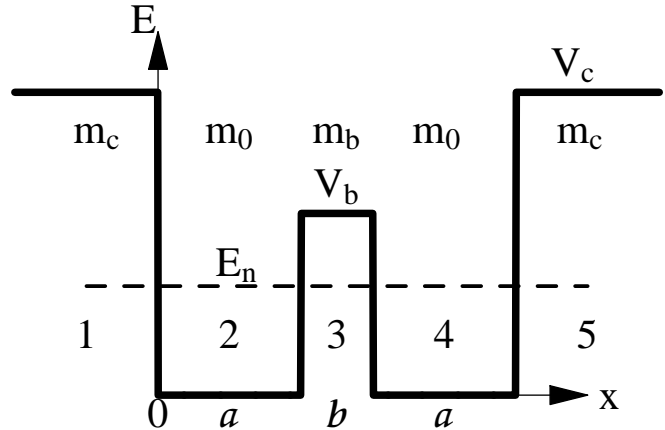}
\caption{\label{fig:DQW} Symmetric double quantum well with
central barrier of width $b$ and height $V_b$. The eigenenergy
$E_n$ is referenced from the bottom of wells of width $a$.
Electron effective mass in regions $1-5$ is assumed to be
different.}
\end{figure}

The one-dimensional DQW with flat potentials in each of
regions $1-5$, as shown in \fref{fig:DQW}, is described by the
following piecewise function of coordinate $x$
\begin{equation}\label{potential}
V(x)=\cases{
 V_c &if $x< 0$ \\
 0 &if $0\le x\le a$\\
 V_b &if $a<x< a+b$\\
 0 &if $a+b\le x\le 2a+b$\\
 V_c &if $x> 2a+b$ \\},
\end{equation}
where $V_c$ is the confining potential (referenced from the bottom
of wells) and $V_b$ is the height of central barrier separating
two identical quantum wells. The mirror symmetry of the system
ensures that the quantum states of such a DQW have either even or
odd parity.

Only bound states will be considered here.  These states can be
normalized to unity over entire $x$ axis. The wave function
$\psi(x)$ in the regions $1-5$ has the following shapes:
\begin{equation}\label{psi}
\eqalign{
 \psi_1=&B_1\rme^{\chi_c x},\\
 \psi_2=&A_1\sin kx+C_1\cos kx, \\
 \psi_3=&B_2\rme^{\chi_b(a-x)}+B_3\rme^{-\chi_b(a+b-x)}, \\
 \psi_4=&A_2\sin k(2a+b-x)+C_2\cos k(2a+b-x), \\
 \psi_5=&B_4\rme^{\chi_c(2a+b-x)},
 }
\end{equation}
where $k$ is the free-electron wave vector,
$k=\sqrt{2m_0E/\hbar^2}$, in the quantum wells of width~$a$. The
energy $E$ is referenced from the bottom of the wells. The wave
vectors of evanescent waves in the exponents are
$\chi_b=\sqrt{(2m_b/\hbar^2)(V_b-E)}$ and
$\chi_c=\sqrt{(2m_c/\hbar^2)(V_c-E)}$, where we have introduced
different electron masses, namely, $m_0$ inside the wells, $m_b$
in the barrier and $m_c$ in the confining potential. This is
typical to semiconductor heterostructures, where the DQW is made
of nanometer layers having different forbidden energy gaps. As a
result, the  electron  effective mass depends on coordinate $x$.

In equations~\eref{psi} there are eight unknown coefficients that
must be calculated. Because of symmetry, the number of
coefficients, in fact, can be reduced. However we shall not do
this since the Gr{\"o}bner basis algorithm  will take account of
symmetry properties of polynomials automatically. The standard
BenDaniel-Duke boundary condition~\cite{BenDaniel66} which takes
into account mass difference on right ($r$) and left ($l$) sides
of the potential step at coordinates~$X=0$, $X=a$, $X=a+b$ and
$X=2a+b$ will be used
 \numparts
\begin{eqnarray}\label{BenDaniel}
\psi_{r}(X^+)&=\psi_{l}(X^-)\, ,\\
\frac{1}{m_r}\frac{\partial\psi_{r}}{\partial
x}\Big\vert_{X^+}&=\frac{1}{m_l}\frac{\partial\psi_{l}}{\partial
x}\Big\vert_{X^-}.
\end{eqnarray}
\endnumparts
Equations~\eref{psi} and the boundary conditions yield the system
of eight linearly dependent equations
 \numparts
\begin{eqnarray}\label{eqs1}
&B_1-C_1=0, \quad -A_1 k/m_0+B_1\chi_c/m_c=0,\\
-&B_2-B_3\rme^{-b\chi_b}+A_1\sin ak+C_1\cos ak =0,\\
&(B_2\chi_b-B_3\chi_b\rme^{-b\chi_b})/m_b+(A_1k\cos ak-C_1k\sin ak)/m_0 =0,\\
&B_3+B_2\rme^{-b\chi_b}-A_2\sin ak-C_2\cos ak =0,\\
&(B_3\chi_b-B_2\chi_b\rme^{-b\chi_b})/m_b+(A_2k\cos ak-C_2k\sin ak)/m_0 =0,\\
-&B_4+C_2=0, \quad -A_2 k/m_0+B_4\chi_c/m_c=0\, .\label{eqs8}
\end{eqnarray}
\endnumparts
The determinant $D$ that follows from this system determines the
spectrum of discrete energy levels of DQW. The symmetry of the
problem ensures the factorization of the determinant
\begin{equation}\label{det}
D=-m_0^{-4}m_c^{-2}m_b^{-2}\rme^{-2b\chi_b}D_sD_a=0\, ,
\end{equation}
where $D_s$ and $D_a$ refer, respectively, to symmetric and
antisymmetric states,
\begin{equation}\label{enSym}
\eqalign{
D_s=&-km_0\big[(\chi_cm_b-\chi_bm_c)+\rme^{b\chi_b}(\chi_cm_b+\chi_bm_c)\big]\cos
ak+ \\
&\big[(k^2m_bm_c+\chi_b\chi_cm_0^2)+\rme^{b\chi_b}(k^2m_bm_c-\chi_b\chi_cm_0^2)\big]\sin
a k\, , }
\end{equation}
\begin{equation}\label{enAsym}
\eqalign{
D_a=&km_0\big[(\chi_cm_b-\chi_bm_c)-\rme^{b\chi_b}(\chi_cm_b+\chi_bm_c)\big]\cos
ak-\\
&\big[(k^2m_bm_c+\chi_b\chi_cm_0^2)-\rme^{b\chi_b}(k^2m_bm_c-\chi_b\chi_cm_0^2)\big]\sin
a k\, . }
\end{equation}
To advance further the transcendental equations $D_s(k)=0$  and
$D_a(k)=0$ which determine, in turn, the spectrum of symmetric and
antisymmetric discrete energy levels have to be solved explicitly.
Unfortunately these transcendental  equation only can be solved by
numerical methods. If DQW parameter values are known, then roots
of \eref{enSym} and \eref{enAsym} define the spectrum of all wave
vectors $k_n$, or equivalently discrete eigenenergies
$E_n=\hbar^2k_n^2/2m_0$ of the DQW, where $n$ is the energy level
index.

In a special case when the DQW heterostructure is fabricated from
two types of nanolayers (labelled $b$ and $0$) we have that
$V_c=V_b$ and $m_c=m_b$. Then $\chi_c=\chi_b$, and the
determinants \eref{enSym} and \eref{enAsym} simplify to
\begin{equation}\label{detsa}
\eqalign{
 D_{s,a}=&-2k\chi_bm_0m_b\rme^{b\chi_b}\cos ak\pm\\
&\big[(k^2m_b^2+\chi_b^2m_0^2)+\rme^{b\chi_b}(k^2m_b^2-\chi_b^2m_0^2)\big]\sin
ak=0\, ,}
\end{equation}
where plus/minus signs correspond to symmetric/antisymmetric states.
When $m_0=m_b$ further simplification is possible
\begin{equation}\label{detsa2}
 D_{s,a}=2\cos{ak}+(\xi-\xi^{-1})\sin(ak)\pm(\xi+\xi^{-1})\sin(ak)\rme^{-\chi
b}=0,
\end{equation}
where now $k=\sqrt{2m_0E}/\hbar$, $\chi=\sqrt{2m_0(V-E)}/\hbar$
and $\xi=\chi/k=\sqrt{(V-E)/E}\,$. Here the plus/minus sign
corresponds to the antisymmetric/symmetric state relative to the
center of the DQW structure, respectively. The
expression~\eref{detsa2} can be found in
references~\cite{Weisbuch87,Bastard84}, where the energy in the
presented formulae is counted from the top of the wells. When the
barrier width $b\rightarrow\infty$, equation~\eref{detsa2} goes
back to the well known formula for an isolated quantum well.

When the particle energy $E$ is larger than the height $V_b$ of
the barrier but smaller than the confining potential, $V_b<E<V_c$,
the particle still remains localized. The only difference is that
in the regions $2-4$ wave function now oscillates, i.e. the
eigenfunctions $\psi(x)$ here are described by trigonometric
functions only. It is easy to see that the above solution at
$E<V_b$ remains valid if we account for hyperbolic functions
properties $\sinh(\rmi\chi_2)=\rmi\sin\chi_2$,
$\cosh(\rmi\chi_2)=\cos\chi_2$ and notice that in this case
$\chi_b$ can be replaced by
$\rmi\chi_b=\rmi\sqrt{(2m_0/\hbar^2)(E-V_b)}\,$.

\section{Eigenfunctions\label{sec:3}}
The coefficients in the wave function~\eref{psi} depend on $k_n$.
Since the spectrum $k_{n}$ (or $E_n=\hbar^2k_n^2/2m_0$) is
determined by roots of the transcendental equations \eref{enSym}
and \eref{enAsym}, one is obliged to solve these equations using
numerical methods. Nonetheless, as we shall see, the
eigenfunctions can be explicitly calculated with the help of
Gr{\"o}bner basis algorithm~\cite{Cox98,Trott04} without any
reference to the roots at all. Roughly speaking, a Gr{\"o}bner
basis for a system of polynomial equations is a different system
of simpler polynomials having the same roots as the original ones.
Calculation of the Gr{\"o}bner basis to some extent resembles
reduction of square matrix to triangular matrix. For further
calculations it is convenient to introduce the following half
angle variables
\begin{equation}\label{xy}
x=\tan(b k/2), \quad y=\tan(a k/2)
\end{equation}
and express sine and cosine functions in \eref{eqs1}--\eref{eqs8} and
\eref{enSym} (or \eref{enAsym} in case of antisymmetric
eigenfunctions) through polynomial variables $x$ and
$y$,
\begin{equation}
\eqalign{
 \sin ak=&\frac{2x}{1+x^2}\, ,\quad\cos
ak=\frac{1-x^2}{1+x^2}\, ,\\
 \sin bk=&\frac{2y}{1+y^2}\, ,\quad\cos
bk=\frac{1-y^2}{1+y^2}\, . }
\end{equation}
Calculating Gr{\"o}bner basis for coefficients $A,B$ and $C$ and
requesting that new variables $x$ and $y$ to be eliminated, the \textit{Mathematica} program
generates basis which consists of 146 polynomials. However, it should be
stressed that the program can find the Gr{\"o}bner basis only if
 the spectrum equation, either \eref{enSym} or
\eref{enAsym}  is appended to the original polynomial
system~\eref{eqs1}--\eref{eqs8}. The following simplest
polynomials were selected for symmetric states
\begin{equation}\label{solSym}
\eqalign{
& A_1=A_2=C_{2s}\frac{\chi_cm_0}{km_c},\\
&B_1=B_4=C_1=C_{2s},\\
&B_2=B_3=\frac{\pm C_{2s}
m_b\rme^{b\chi_b}\sqrt{k^2m_c^2+\chi_c^2m_0^2}}{m_c\big[k^2m_b^2(1+\rme^{b\chi_b})^2
+\chi_b^2m_0^2(-1+\rme^{b\chi_b})^2\big]^{1/2}}\, , }
\end{equation}
where $C_2$ was replaced by $C_{2s}$ to identify the state
symmetry.  The choice of sign of $B_2$ and $B_3$ coefficients has
to ensure derivative continuity at points $a$ and $a+b$. It is
straightforward to check that the solution \eref{solSym} indeed
satisfies the initial equations~\eref{eqs1}--\eref{eqs8}. In
\eref{solSym} all amplitudes are expressed through a single
coefficient~$C_{2s}$, which in turn can be found from the
normalization condition of the total wave function $\psi(x)$. The
integration over $x$ axis gives the normalization constant in the
form
\begin{equation}
C_{2s}=km_c(G_1+G_2)^{-1/2}
\end{equation}
where
\begin{equation}
G_1=\chi_c^{-1}\big[k^2m_c^2(1+a\chi_c)+m_0\chi_c^2(m_c+m_0a\chi_c)\big]\,
,
\end{equation}
\begin{equation}
G_2=\frac{m_b(k^2m_c^2+\chi_c^2m_0^2)\big[b\chi_bk^2m_b+(k^2m_b+\chi_b^2m_0)\sinh
b\chi_b\big]}
 {\chi_b\big[k^2m_b^2-\chi_b^2m_0^2+(k^2m_b^2+\chi_b^2m_0^2)\cosh
b\chi_b\big]}\, .
\end{equation}
If  all masses are assumed to be equal ($m_0=m_b=m_c=1$) the
normalization constant simplifies to
\begin{equation}
C_{2s}=\sqrt{\chi_b}\Big[\Big(1+\frac{\chi_c^2}{k^2}\Big)
\Big(a\chi_b+\frac{\chi_b^2}{\chi_c^2}+\frac{b\chi_bk^2+(k^2+\chi_b^2)\sinh
b\chi_b}{k^2-\chi_b^2+(k^2+\chi_b^2)\cosh
b\chi_b}\Big)\Big]^{-1/2}.
\end{equation}

Quite similar calculation for antisymmetric $(C_2\rightarrow C_{2a})$
states yields
\begin{equation}\label{solAsym}
\eqalign{
&-B_1=B_4=-C_1=C_{2a},\quad -A_1=A_2=C_{2a}\frac{\chi_cm_0}{km_c},\\
&-B_2=B_3=\frac{\pm
C_{2a}m_b\rme^{b\chi_b}(k^2m_c^2+\chi_c^2m_0^2)^{1/2}}
{m_c\big[k^2m_b^2(-1+\rme^{b\chi_b})^2
+\chi_b^2m_0^2(1+\rme^{b\chi_b})^2\big]^{1/2}}\, , }
\end{equation}
where the choice of sign again follows from the derivative continuity condition.
The normalization constant in this case is
\begin{equation}
C_{2a}=km_c(H_1+H_2)^{-1/2}\, ,
\end{equation}
where
\begin{equation}
H_1=\chi_c^{-1}\big[k^2m_c^2(1+a\chi_c)+m_0\chi_c^2(m_c+m_0a\chi_c)\big]\,
,
\end{equation}
\begin{equation}
H_2=\frac{m_b(k^2m_c^2+\chi_c^2m_0^2)\big[-b\chi_bk^2m_b+(k^2m_b+\chi_b^2m_0)\sinh
b\chi_b\big]}
 {\chi_b\big[-k^2m_b^2+\chi_b^2m_0^2+(k^2m_b^2+\chi_b^2m_0^2)\cosh
b\chi_b\big]}\, .
\end{equation}
When all masses becomes equal the normalization constant
$C_{2a}$ reduces to
\begin{equation}
C_{2a}=\sqrt{\chi_b}\Big[\Big(1+\frac{\chi_c^2}{k^2}\Big)
\Big(a\chi_b+\frac{\chi_b^2}{\chi_c^2}+\frac{-b\chi_bk^2+(k^2+\chi_b^2)\sinh
b\chi_b}{-k^2+\chi_b^2+(k^2+\chi_b^2)\cosh
b\chi_b}\Big)\Big]^{-1/2}.
\end{equation}
As far as a more general non symmetric DQW problem concerns, the
calculations of the  Gr{\"o}bner basis indicates that, in contrast
to solutions \eref{solSym} and \eref{solAsym}, at least one of the
coefficients $A$, $B$, or $C$ includes the trigonometric
functions. In this case the determinant $D$ does not factorize to
symmetric and asymmetric parts either.

\section{Dipole matrix element\label{sec:4}}

The knowledge of eigenfunctions allows one to carry on with
analytical calculations. As an example we shall find closed form
expression for dipole matrix elements between even $\psi_s(k_n,x)$
and odd $\psi_a(k_m,x)$ discrete states
\begin{equation}
d_{ns,ma}=\int_{-\infty}^{+\infty}
\psi_s^*(k_n,x)\Big(x-a-\frac{b}{2}\Big)\psi_a(k_m,x)\rmd x=2d_1+2d_2+d_3\, .
\end{equation}
Here the subscripts $s$ and $a$ refer to, respectively, even and
odd symmetry states and $d_i$~is the contribution of the $i$-th
region indicated in the~\fref{fig:DQW}. For a general case the
expressions for dipole components $d_{ns,ma}$ are rather
complicated~\cite{notebook}. For simplicity below we present the
expressions for the case when masses in all regions are equal,
$m_c=m_b=m_0$ and the central and confining barrier heights
coincide, $\chi_c=\chi_b=\chi$. Since the energy of symmetric and
antisymmetric states differ the wave vectors $k$ and $\chi$ are
supplied by indices $s$ and $a$. Thus the dipole expression have
two kind of the wave vectors $k_s$ and $k_a$, and evanescent modes
$\chi_s$ and $\chi_a$.

\begin{figure}[t]
\centering
\includegraphics[width=8cm]{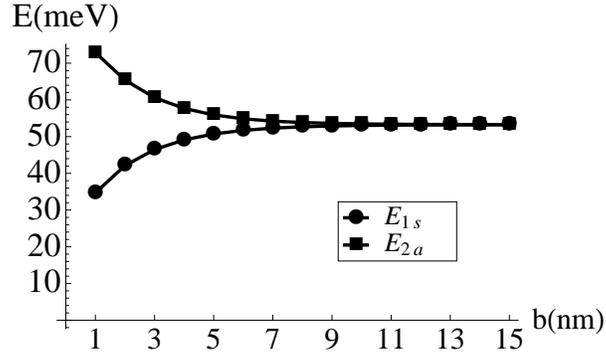}
\caption{\label{fig:2} The energies of
GaAs/Ga$_{0.8}$Al$_{0.2}$As DQW as a functions of the central
barrier width at $a=6~\textrm{nm}$.}
\end{figure}

\begin{figure}[t]
\centering
\includegraphics[width=8cm]{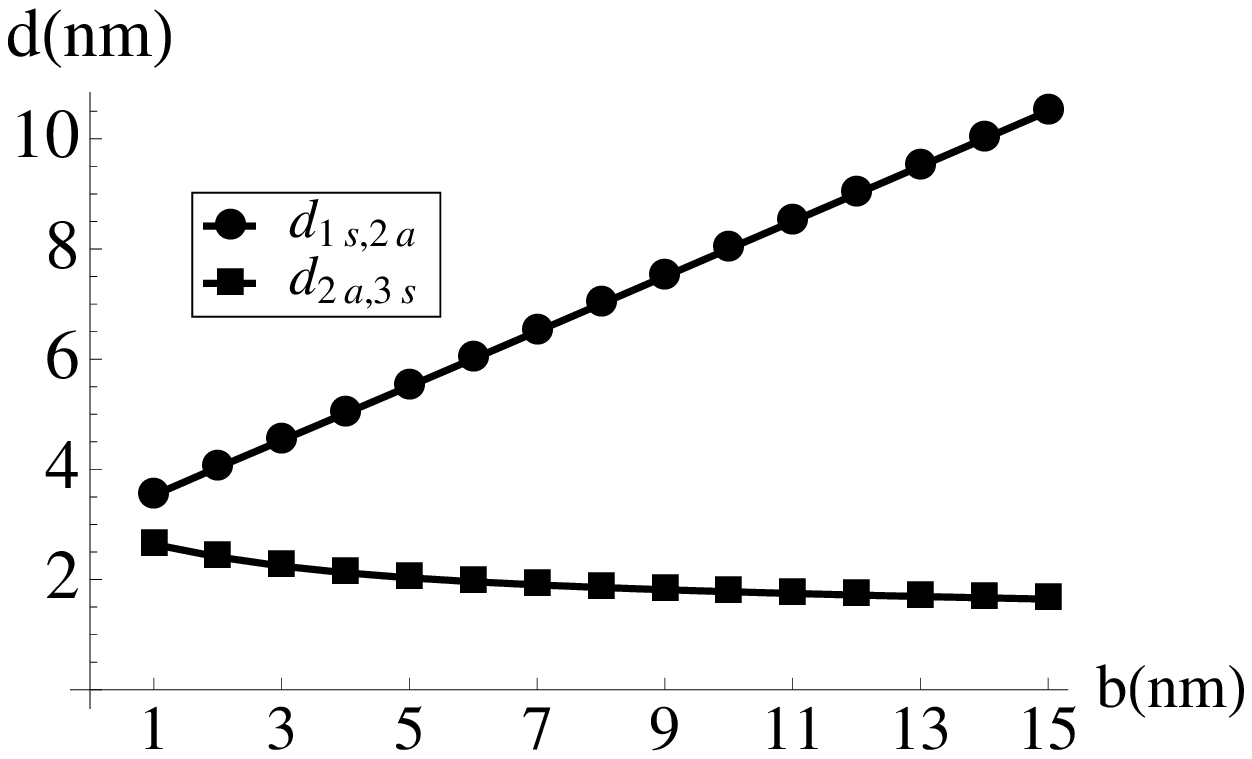}a)
\includegraphics[width=8cm]{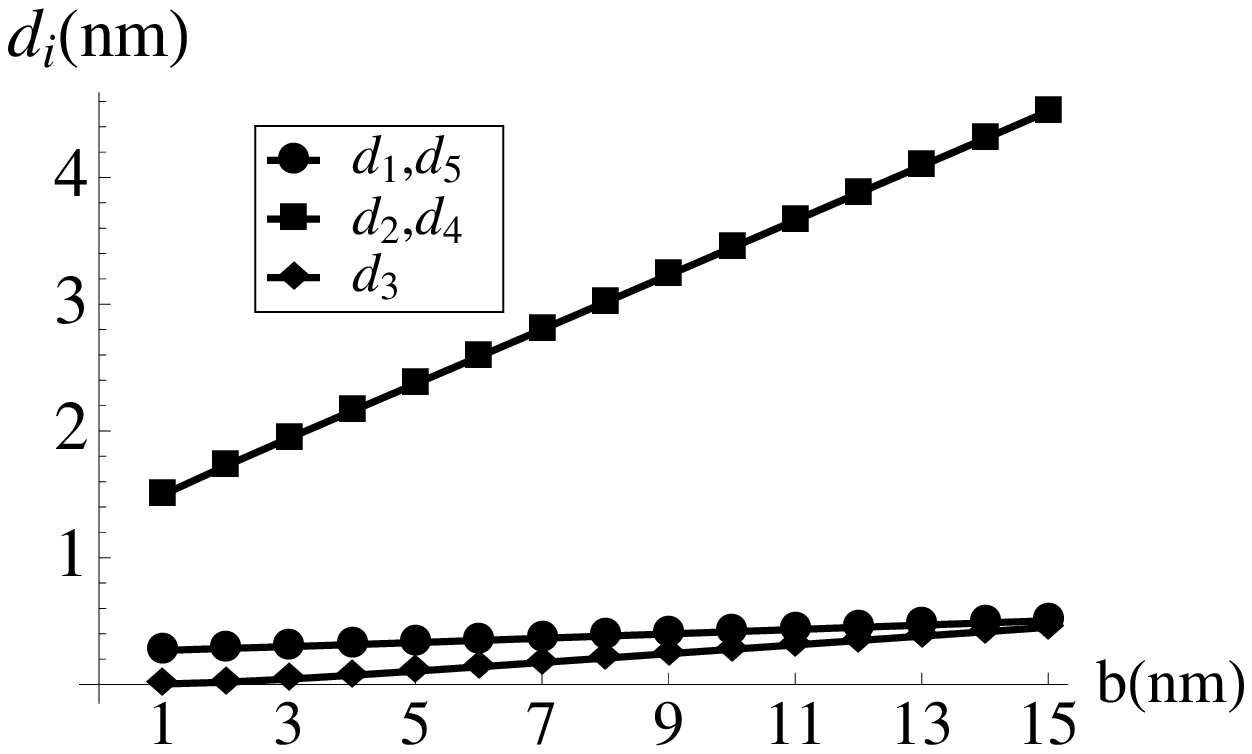}b)
\caption{\label{fig:3} a)~Dipole matrix elements $d_{1s,2a}$  and
$d_{2a,3s}$ as a function of barrier width~$b$. b)~Contribution of
individual regions to dipole matrix $d_{1s,2a}$.}
\end{figure}

In the first and fifth regions the contribution to dipole is
\begin{equation}
d_1=d_5=d_{1N}/d_{1D}\, ,
\end{equation}
where
\begin{eqnarray}
d_{1N}&=&
 k_sk_a\big[2+(2a+b)(\chi_s+\chi_a)\big]r_sr_a\, ,\\
 r_s&=&\sqrt{\chi_s\big[(k_s^2-\chi_s^2)+(k_s^2+\chi_s^2)\cosh b\chi_s\big]}\, ,\nonumber\\
 r_a&=&\sqrt{\chi_a\big[-(k_a^2-\chi_a^2)+(k_a^2+\chi_a^2)\cosh
 b\chi_a\big]}\, ,\nonumber
\end{eqnarray}
and
\begin{eqnarray}
d_{1D}&=& 2(\chi_s+\chi_a)^2\sqrt{(k_s^2+\chi_s^2)(k_a^2+\chi_a^2)}\:\delta_s\delta_a\, ,\\
\delta_s&=&\Big(-\chi_s^2(1+a\chi_s)+k_s^2\big[1+(a+b)\chi_s\big]+\nonumber \\
& &(k_s^2+\chi_s^2)\big[(1+a\chi_s)\cosh b\chi_s+\sinh
b\chi_s\big]\Big)^{1/2} ,\nonumber\\
\delta_a&=&\Big(\chi_a^2(1+a\chi_a)-k_a^2\big[1+(a+b)\chi_a\big]+\nonumber\nonumber \\
& &(k_a^2+\chi_a^2)\big[(1+a\chi_a)\cosh b\chi_a+\sinh
b\chi_a\big]\Big)^{1/2} .\nonumber
\end{eqnarray}

In the second region it is
\begin{equation}
d_2=d_{2N}/d_{2D}\, ,
\end{equation}
where
\begin{eqnarray}
d_{2N}&=& -\frac{1}{2}r_sr_a\big[(k_s-k_a)^2(p_1-p_3)-(k_s+k_a)^2(p_2-p_4)\big]\, ,\\
p_1&=&\big[b(k_s+k_a)(k_s\chi_a+k_a\chi_s)+2k_sk_a-2\chi_s\chi_a\big]\cos a(k_s+k_a)\, ,\nonumber\\
p_2&=&\big[b(k_s-k_a)(k_s\chi_a-k_a\chi_s)-2k_sk_a-2\chi_s\chi_a\big]\cos a(k_s-k_a)\, ,\nonumber\\
p_3&=&\big[b(k_s+k_a)(k_sk_a-\chi_s\chi_a)-2k_s\chi_a-2k_a\chi_s\big]\sin a(k_s+k_a)\, ,\nonumber\\
p_4&=&\big[b(k_a-k_s)(k_sk_a+\chi_s\chi_a)+2k_a\chi_s-2k_s\chi_a\big]\sin
a(k_s-k_a)\, ,\nonumber
\end{eqnarray}
\begin{equation}
d_{2D}=\frac{(k_s^2-k_a^2)^2}{(\chi_s+\chi_a)^2}d_{1D}\, .
\end{equation}
One can see that trigonometric functions, which will give
oscillations of matrix elements vs. the well width, appear only
here.

The third (barrier) region contribution to dipole is
\begin{equation}
d_3=d_{3N}/d_{3D}\, ,
\end{equation}
where
\begin{eqnarray}
d_{3N}&=& -4\rme^{\frac{1}{2}b(\chi_s+\chi_a)}k_sk_ar_sr_a\big[v_1\cosh\frac{b\chi_a}{2}+v_2\sinh\frac{b\chi_a}{2}\big]\, ,\\
v_1 &=&-b\chi_a (\chi_s^2-\chi_a^2)\cosh\frac{b\chi_s}{2}+4\chi_s\chi_a \sinh\frac{b\chi_s}{2}\, ,\nonumber\\
v_2
&=&-2(\chi_a^2+\chi_2^2)\cosh\frac{b\chi_s}{2}+4b\chi_s(\chi_s^2-\chi_a^2)\sinh\frac{b\chi_s}{2}\,\nonumber
,
\end{eqnarray}
and
\begin{eqnarray}
d_{3D}&=&\frac{1}{2}(\chi_s-\chi_a)^2s_ss_ad_{1D}\, ,\\
s_s &=&\sqrt{\frac{(1+\rme^{b\chi_s})^2k_s^2+(-1+\rme^{b\chi_s})^2\chi_s^2}{k_s^2+\chi_s^2}}\, ,\nonumber\\
s_a
&=&\sqrt{\frac{(-1+\rme^{b\chi_a})^2k_a^2+(1+\rme^{b\chi_a})^2\chi_a^2}{k_a^2+\chi_a^2}}\,\nonumber
.
\end{eqnarray}

\Fref{fig:2} shows the dependencies of the first two energy levels
$E_{1s}$ and $E_{2a}$ as a function of the inner barrier width.
The following parameter values that are typical to
GaAs/Ga$_{0.8}$Al$_{0.2}$As DQW heterostructures, were used for
production of pictures: $a=6~\textrm{nm}$, $b=(1-15)~\textrm{nm}$,
$V_c=V_b=0.1671~\textrm{eV}$, $m_0=0.067m_e$, $m_c=m_b=0.0836m_e$,
where $m_e$ is the electron mass in the vacuum. The increase of
energy difference between levels with the decrease of $b$ is
assigned to tunnel coupling of levels.

\Fref{fig:3}a demonstrates, respectively, the size of optical
dipole matrix elements between a pairs of adjacent levels,
$d_{1s,2a}$ and $d_{2a,3s}$, as a function of barrier width.
\Fref{fig:3}b shows the contribution of individual regions to the
dipole $d_{1s,2a}$.   It is clear that a general trend and
magnitude of dipole elements in \fref{fig:3}a can be understood if
one assumes that only quantum wells contribute to the total
dipole. In this approximation  the functions
$\psi_1=\psi_3=\psi_5=0$ while the $\psi_2$ and $\psi_4$ can be
approximated by half-period sine functions. Then $d_{1s,2a}$
reduces to
\begin{equation}
d_{1s,2a}\approx\frac{2}{a}\int_0^a \sin \frac{\pi
x}{a}\Bigl(x-a-\frac{b}{2}\Bigr)\bigl(-\sin \frac{\pi
x}{a}\bigr)\mathrm{d}x=\frac{a+b}{2}.
\end{equation}
The formula shows that dipole  size increases linearly with the
barrier width $b$  as long as $b$ remains much smaller than
exciting light period. For $2a-3s$ optical transitions one of
sines should be replaced by $\sin (2\pi x/a)$. Then, similar
calculation yields $d_{2a,3s}\approx 16a/9\pi^2$, which is
independent of barrier width. The deviations from the obtained
expressions in \fref{fig:3}a come from the evanescent mode
contribution in barrier and confining potential regions.

In conclusion, the presented example shows that application of
Gr{\"o}bner  basis algorithm in some cases allows to find closed
form expressions for the total wave function and, therefore, to
calculate the dipole matrix elements exactly without directly
solving the transcendental equations that determines the spectrum
of the DQW. Of course, the described method can be applied to
other quantum systems for which eigenvalue equations cannot be
explicitly solved as well.

\ack This work was supported by EU grant ''Science for Business
and Society'' No:~VP2-1.4-{\= U}M-03-K-01-019


\end{document}